\documentstyle[aps,prl,floats,epsf]{revtex}
\tighten
\draft
\begin{document}

\twocolumn[\hsize\textwidth\columnwidth\hsize\csname
@twocolumnfalse\endcsname 

\title{On the Normalization of the Neutrino-Deuteron Cross Section}
\author{\mbox{John F. Beacom$^{1}$} and \mbox{Stephen J. Parke$^{2}$}}
\address{\mbox{$^1$ NASA/Fermilab Astrophysics Center, 
Fermi National Accelerator Laboratory, Batavia, Illinois 60510-0500}
\mbox{$^2$ Theoretical Physics Department, Fermi National 
Accelerator Laboratory, Batavia, Illinois 60510-0500}
\\{\tt beacom@fnal.gov}, {\tt parke@fnal.gov}}
\date{June 12, 2001}
\maketitle

\begin{abstract}
As is well-known, comparison of the solar neutrino fluxes measured in
SuperKamiokande (SK) by $\nu + e^- \rightarrow \nu + e^-$ and in the
Sudbury Neutrino Observatory (SNO) by $\nu_e + d \rightarrow e^- + p +
p$ can provide a ``smoking gun'' signature for neutrino oscillations
as the solution to the solar neutrino puzzle.  This occurs because SK
has some sensitivity to all active neutrino flavors whereas SNO can
isolate electron neutrinos.  This comparison depends crucially on the
normalization and uncertainty of the theoretical charged-current
neutrino-deuteron cross section.  We address a number of
effects which are significant enough to change the interpretation of
the SK--SNO comparison.
\end{abstract}

\vspace{0.25cm}]
\narrowtext


Both SK and SNO are sensitive to solar neutrinos with energies above
about 5 MeV.  In SK, these are detected by $\nu + e^- \rightarrow \nu
+ e^-$, with possible (indistinguishable) contributions from all
active flavors.  In particular, if there are $\nu_e \rightarrow
\nu_\mu, \nu_\tau$ oscillations, then the latter contribute to the
measured flux with a cross section 6--7 times smaller than for
$\nu_e$.  In SNO, on the other hand, the detection reaction $\nu_e + d
\rightarrow e^- + p + p$ can isolate the $\nu_e$ flux.

The measured flux of solar neutrinos in SK, in units of the expected
electron neutrino flux from the Standard Solar Model
(SSM)~\cite{BPB2000}, is $0.45 \pm 0.02$, with the systematic
uncertainty dominating~\cite{SK1258}.  This measured flux may or may
not include a contribution from $\nu_\mu$, $\nu_\tau$ (in any linear
combination, since they interact via the neutral current).  In an
energy-independent (as suggested by the absence of any distortion in
the SK recoil electron spectrum~\cite{SK1258}) two-flavor oscillation
scenario, there are two extreme cases~\cite{SKSNOtheory}.  First, for
$\nu_e \rightarrow \nu_s$, the $\nu_e$ flux in these units is 0.45,
and the undetectable $\nu_s$ (sterile neutrino) flux is 0.55.  Second,
for $\nu_e \rightarrow \nu_\mu, \nu_\tau$, the $\nu_e$ flux is 0.34,
and the $\nu_\mu$, $\nu_\tau$ flux 0.66, so that the measured flux in
SK is $0.34 + 0.66/6 = 0.45$.  In the first case, SNO will measure
0.45, and in the second case, 0.34.  More generally, these arguments
can be rephrased as a ratio to eliminate the SSM flux normalization
and its $\simeq 20\%$ uncertainty, since the total incident flux is
the same for both SK and SNO.  Also, a small correction is necessary
for the $hep$ neutrinos~\cite{BPB2000,Marcucci} that add to the
dominant $^8$B neutrinos.

This possible difference of 0.11 is small enough that the
uncertainties must be scrutinized closely.  Below, we closely follow
the analogous results for the percent-level corrections to the
theoretical $\bar{\nu}_e + p \rightarrow e^+ + n$ cross
section~\cite{invbeta1,invbeta2} that are necessary to achieve
agreement with experiment~\cite{Declais}.

While results from SNO~\cite{SNOproposal,SNOdetector} have not yet
been reported, they have said at conferences~\cite{SNOtalks} that they
expect their flux measurement uncertainty to be dominated by the $\sim
3\%$ theoretical uncertainty~\cite{BCK,NSGK} on the neutrino-deuteron
cross section, at least eventually.

In the SNO proposal~\cite{SNOproposal}, the uncertainty on the
neutrino-deuteron cross sections was assumed to be about 10\% (for
comparison, the experimental measurements of neutrino-deuteron cross
sections have uncertainties of 10 -- 40\%~\cite{sigmanud}).  Since
that time, the calculations have been redone by a number of authors,
with decreasing quoted uncertainties.  The most refined recent
calculations are those of Butler, Chen, and Kong (BCK)~\cite{BCK} and
Nakamura, Sato, Gudkov and Kubodera (NSGK)~\cite{NSGK}.  Each claims
an uncertainty of about 3\% in the range of energies relevant for
solar neutrinos, and they agree to about 1\%~\cite{BCK}.  It is now
important to consider issues that affect the normalization of the
total cross section at a comparable level.  Three such effects have
been overlooked by BCK and NSGK; these effects are comparable and add
constructively.

First, at low energies, the neutrino-deuteron cross section is
dominated by the Gamow-Teller transition, so that the cross section
scales as $g_A^2$, where $g_A$ is the weak axial coupling to nucleons,
and the angular distribution~\cite{invbeta2} of the outgoing electron
is nearly of the form $1 - \frac{1}{3} \cos\theta$.  The present value
of $g_A$ is $-1.267 \pm 0.004$~\cite{RPP}.  BCK use $-1.26$, which
makes their cross section about 1\% too small, and NSGK use $-1.254$,
which makes their cross section about 2\% too small.  This effect is
trivial in nature but must be included.

Second, a more subtle effect occurs because both BCK and NSGK use the
Fermi constant $G_F$ as determined from muon decay~\cite{RPP}.  The
radiative corrections to low-energy (much less than the $W$ mass) weak
processes are frequently divided into ``inner'' (which are
energy-independent) and ``outer'' (which are generally
energy-dependent) corrections.  The inner radiative correction is
universal to a given reaction and those related to it by crossing
symmetry, and can thus be considered to renormalize $G_F$ for each set
of diagrams.  This renormalization is different for purely leptonic
and semileptonic processes, e.g., muon decay and neutron beta decay
(it is also different for neutral-current weak processes involving a
nucleon).  Since the inner radiative correction arises from diagrams
with internal $\gamma$ and $Z$ exchanges with high momenta, the quark
structure of the nucleon is resolved and the fact that the nucleon is
bound in a deuteron is irrelevant.  Thus the inner radiative
correction increases the $\nu_e + d \rightarrow e^- + p + p$ cross
section by 2.4\%, just as for $\bar{\nu}_e + p \rightarrow e^+ +
n$~\cite{invbeta2}.

Third, the outer radiative correction must be considered.  This arises
from bremsstrahlung diagrams with a single external photon and from
diagrams with low-momentum internal $\gamma$ exchange.  The only
calculation of the radiative corrections for $\nu_e + d \rightarrow
e^- + p + p$ is given by Towner~\cite{Towner}.  If the bremsstrahlung
photon energy is included in the deposited energy, the total radiative
corrections are only mildly energy-dependent, and are given in
Towner's Table II as about 4.4\%.  Taking this at face value, and
subtracting the above inner radiative correction of 2.4\%, Towner's
result for the outer radiative correction to $\nu_e + d \rightarrow
e^- + p + p$ is nearly constant at about 2\%.  This correction is
somewhat larger than the corresponding corrections for $\bar{\nu}_e +
p \rightarrow e^+ + n$~\cite{invbeta1}, and $\nu_e + n \rightarrow e^-
+ p$~\cite{Dicus}, each about 1\% and decreasing with increasing
energy.  In all cases, the total cross section is increased.

Thus, BCK and NSGK each underestimate the total cross section by about
6\%.  If the overlooked normalization effects discussed above are
taken into account, it does seem reasonable to use the quoted nuclear-physics
uncertainty of 3\%.  It should be noted that these very sophisticated
calculations differ from the simplest treatment only by about 10\%
(while a number of corrections contribute, none is as large as their
sum)~\cite{BCK,NSGK}.  Thus, roughly speaking, in order for the cross
section to be known to 3\%, the corrections only have to be known to
30\%, which seems reasonable.

These simple considerations are backed up by the detailed results of
BCK and NSGK.  BCK have shown that up to next-to-next-to-leading order
in their effective field theory treatment, only one unmeasured
parameter $L_{1,A}$ (it appears at next-to-leading order) appreciably
affects their result.  They have also shown that their effective field
theory series is convergent, with the contribution from each order
about ten times smaller than the previous order.  NSGK state that
Ref.~\cite{YHH} did not include all of the known exchange-current
contributions; when also neglected in NSGK, NSGK agree with
Ref.~\cite{YHH} to about 1\%.  Similarly, BCK can reproduce the
results of Ref.~\cite{YHH} by adjusting the value of $L_{1,A}$.  BCK
and NSGK agree with each other and Ref.~\cite{KN} at the 1--2\% level.

Bahcall, Krastev, and Smirnov~\cite{BKS} quote a theoretical
uncertainty of 6\% on the neutrino-deuteron total cross section based
on nuclear-physics differences in the calculated cross sections of
Refs.~\cite{YHH,KN,BL}.  Since then, almost all of the difference
between Refs.~\cite{YHH,KN} has been explained by NSGK.  Similarly for
the difference between Ref.~\cite{KN} and the effective range
calculation of Ref.~\cite{BL}, where no exchange current effects are 
included.  Thus, a theoretical uncertainty of 6\% seems too conservative.

\begin{figure}[t]
\centerline{\epsfxsize=3.25in \epsfbox{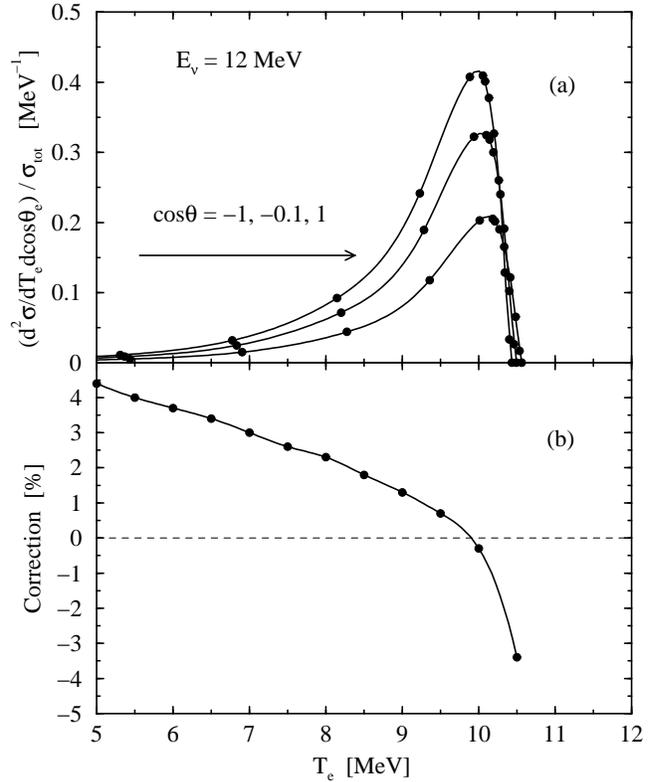}}
\caption{{\bf (a)}: The differential cross section without radiative
corrections for $\nu_e + d \rightarrow e^- + p + p$ as a function of
the electron kinetic energy $T_e$, for selected values of the electron
cosine $\cos\theta_e$, for $E_\nu = 12$ MeV.  The points are from a
table provided by Kubodera (private communication), based on
Ref.~\protect\cite{KN}, and the lines are spline fits.  The height
differences for different $\cos\theta_e$ are mostly accounted for by
the angular distribution $d\sigma/d\cos\theta_e \simeq 1 - \frac{1}{3}
\cos\theta_e$ (note $\langle \cos\theta \rangle \simeq -0.1$).  {\bf
(b)}: Total radiative corrections $\frac{\alpha}{\pi}
g_{I}(E_\nu,T_e)$ to the differential cross section as a function of
$T_e$, using Towner's Table I (points)~\protect\cite{Towner}, also for
$E_\nu = 12$ MeV.  The line is a spline fit.  For the curves in both
panels, other neutrino energies yield similar shapes when considered
as a function of $T_e/E_\nu$.}
\end{figure}

So far, we have only considered the corrections to the normalization
of the total cross section.  We now turn to the radiative corrections
to the differential cross section.  The differential cross section
without radiative corrections for a representative neutrino energy is
shown in Fig.~1(a).  SNO may initially operate with a high electron
energy threshold, above the peak in the electron spectrum expected
from the $^8$B neutrino spectrum.  If so, a slight downward shift (of
order 100 keV) in the electron energies can cause a few-percent
decrease in the number of events above threshold (when integrated over
the $^8$B spectrum).  For photons below about 1 MeV, the
Compton-scattered electrons will not produce detectable \v{C}erenkov
light (H. Robertson, private communication).  Thus, one of the effects
of bremsstrahlung will be to lower the electron energies from the case
with no bremsstrahlung.  In Towner's Table I, the bremsstrahlung
energy is considered undetected, and the correction to $d\sigma/dT_e$
is as large as $+5\%$ to $-4\%$ from low to high scattered electron
energies, and nearly vanishing at electron kinetic energy $T_e \simeq
E_\nu - 2$ MeV.  This is shown in Fig.~1(b).  The qualitative features
of Towner's total radiative corrections can be reproduced by a small
downward shift in the electron energies.

Using Towner's Table I, where the bremsstrahlung energy is considered
undetected, we can calculate the total cross section by integrating
\begin{equation}
\frac{d\sigma}{dT_e} = \left(\frac{d\sigma}{dT_e}\right)^0
\times \left[1 + \frac{\alpha}{\pi} g_{I}(E_\nu,T_e)\right]\,,
\end{equation}
where the superscript 0 indicates the cross section without radiative
corrections.  Using Towner's Table II, where the bremsstrahlung energy
is considered to add to the detected electron energy, we can calculate
the total cross section by integrating
\begin{equation}
\frac{d\sigma}{dX} = \left(\frac{d\sigma}{dX}\right)^0
\times \left[1 + \frac{\alpha}{\pi} g_{II}(E_\nu,X)\right]\,,
\end{equation}
where $X$ is the sum of electron and bremsstrahlung energies (without
radiative corrections, this is just the electron energy).  In the
latter case, the total radiative corrections are nearly constant at
$\frac{\alpha}{\pi} g_{II}(E_\nu,X) \simeq 4.4\%$.  If there are no
cuts on the kinematic variables, then these two integrals must be
identical.  We explicitly made this test (using spline fits) for
$E_\nu = 12$ MeV, and found it not to be the case.  In the first case,
we find an average correction of 0.7\% above $T_e = 5$ MeV (this is
also obvious by inspection of Fig.~1, if the correction is evaluated
at the average energy for the differential cross section), much less
than the 4.4\% for the second case.  The neglected fraction of the
$d\sigma/dT_e$ integral below 5 MeV is about 3\%, so in order to
reproduce the integrated total corrections of 4.4\%, the correction
below 5 MeV (not given in Towner's Table I) would have to be of order
100 times larger than that above 5 MeV.  Thus, we are unable to see
how the results of Towner's Table I can be consistent with the
seemingly reasonable results of his Table II.

Given the importance of the radiative corrections for the SK--SNO
comparison, additional work is needed, in particular on the
bremsstrahlung spectrum.  For a sufficiently soft bremsstrahlung
spectrum, the corrections to the total cross section will be
applicable.  Otherwise, corrections to the differential cross section
will also be necessary.

In conclusion, three overlooked effects conspire to increase the
normalization of the total cross section for $\nu_e + d \rightarrow
e^- + p + p$ by about 6\%.  As noted, the uncertainty in the measured
neutrino flux in SNO is expected to be eventually dominated by the
uncertainty in the theoretical cross section.  In addition, if SNO
operates with a high threshold, the effects of the radiative
corrections on the differential cross section must be considered.
These effects, if not taken into account, could qualitatively change
the outcome of the SK--SNO comparison, which is a SSM-independent test
for the appearance of the active flavors $\nu_\mu$, $\nu_\tau$
resulting from neutrino oscillations (e.g., in a
$\phi_{\nu_\mu,\nu_\tau}/\phi_{SSM}$ versus $\phi_{\nu_e}/\phi_{SSM}$
plot, where $\phi$ is the neutrino flux).  If the effects discussed
above, in particular the QED radiative corrections, are correctly
taken into account, then the 3\% theoretical uncertainty indicated by
BCK and NSGK for the neutrino-deuteron cross section in the energy
range appropriate for solar neutrinos is attainable.

We thank Malcolm Butler, Jiunn-Wei Chen, Rocky Kolb, Kuniharu
Kubodera, Ian Towner, Petr Vogel, and especially Hamish Robertson for
discussions.  J.F.B. was supported by a David N. Schramm Fellowship
(funded in part by NASA under NAG5-7092).  Fermilab is operated by URA
under DOE contract No. DE-AC02-76CH03000.



\end{document}